\documentclass[aps,pra,twocolumn,showpacs,amsmath,amssymb]{revtex4}

\usepackage{graphicx} 
\usepackage{dcolumn} 
\usepackage{bm} 
\usepackage{epsfig}
\usepackage{amsmath}
\usepackage{amssymb}
\usepackage{hyperref}
\usepackage{float}
\usepackage{color}
\usepackage{ulem}
\usepackage{slashed} 
\begin{document}
\newcommand{\bsy}[1]{\mbox{${\boldsymbol #1}$}} 
\newcommand{\bvecsy}[1]{\mbox{$\vec{\boldsymbol #1}$}} 
\newcommand{\bvec}[1]{\mbox{$\vec{\mathbf #1}$}} 
\newcommand{\btensorsy}[1]{\mbox{$\tensor{\boldsymbol #1}$}} 
\newcommand{\btensor}[1]{\mbox{$\tensor{\mathbf #1}$}} 
\newcommand{\tensorId}{\mbox{$\tensor{\mathbb{\mathbf I}}$}} 
\newcommand{\be}{\begin{equation}}
\newcommand{\ee}{\end{equation}}
\newcommand{\bea}{\begin{eqnarray}}
\newcommand{\eea}{\end{eqnarray}}
\newcommand{\e}{\mathrm{e}}
\newcommand{\arccot}{\mathrm{arccot}}
\newcommand{\arctanh}{\mathrm{arctanh}}

\title{Negative refraction in the relativistic electron gas}
 
\author{C. A. A. de Carvalho$^{1,2}$, D. M. Reis$^{1}$, and D. Szilard$^{2}$}

\affiliation{
$^1$Centro Brasileiro de Pesquisas F\'{\i}sicas - CBPF, Rua Dr. Xavier Sigaud 150, Rio de Janeiro - RJ, 22290-180, Brazil\\
$^2$Instituto de F\'{\i}sica, Universidade Federal do Rio de Janeiro - UFRJ, Caixa Postal  68528, Rio de Janeiro - RJ, 21945-972, Brazil}

\begin{abstract}
We show that a gas of relativistic electrons is a left-handed material at low frequencies by computing the effective electric permittivity and effective magnetic permeability that appear in Maxwell's equations in terms of the responses appearing in the constitutive relations, and showing that the former are both negative below the {\it same} frequency, which coincides with the zero-momentum frequency of longitudinal plasmons. We also show,  by explicit computation, that the photonic mode of the electromagnetic radiation does not dissipate energy, confirming that it propagates in the gas with the speed of light in vacuum, and that the medium is transparent to it. We then combine those results to show that the gas has a negative effective index of refraction $n_{\rm eff}=-1$. We illustrate the consequences of this fact for Snell's law, and for the reflection and transmission coefficients of the gas. 
\end{abstract}


\date{\today}

\maketitle

\section{Introduction}

More than sixty years ago, Viktor Veselago \cite{Veselago} studied the electrodynamics of materials with electric permittivity $\epsilon$ and magnetic permeability $\mu$ simultaneously negative. At the time, the study was just a theoretical exercise, since no examples of  materials exhibiting those characteristics were known. Years later, it was even claimed that those materials were not ``naturally occurring" \cite{Pendry}, because negative magnetic responses did not seem to occur in Nature. 

In fact, materials with both $\epsilon$ and $\mu$ negative were eventually {\it constructed}, thanks to the engineering and assembling of special nanostructures called split-ring resonators - SRR's \cite{Smith}. Such structures amplified magnetic responses, making them comparable to electric ones, and could be arranged so as to yield negative values for both $\epsilon$ and $\mu$. Those materials are called left-handed (LHM's), because electromagnetic (EM) waves propagate in them with a Poynting vector opposite to the wavevector.

In a series of articles \cite{AragaoPRDS2016, ReyezEPL2016, JPP, ReisAdP2018}, we have studied the relativistic electron gas (REG) as a candidate to realize, in a {\sl natural} way (we will clarify what we mean by this later on), the physical situation proposed by Veselago as a mere theoretical exercise. There, we have derived the constituent relations for the gas, which involved various electric and magnetic responses, and showed that some of them could be negative below certain frequencies, and that there were regions where electric and magnetic responses could be simultaneously negative. 

However, the quantities that Veselago assumed to be simultaneously negative were the electric permittivity and magnetic permeability appearing in Maxwell's equations. Although in nonrelativistic systems these are the same quantities that appear in the constitutive equations, this is NOT the case for the REG. In fact, we will show that what appears in Maxwell's equations for the REG are combinations of the responses present in the constitutive relations. Those combinations are to be interpreted as the effective electric permittivity and permeability that would have to be simultaneously negative in some frequency range in order to realize the Veselago system. Here, we will construct them, and show that both become negative below the {\it same} frequency value, which we associate with a longitudinal plasmon frequency.

In one of the articles \cite{ReisAdP2018}, from the study of the poles in the EM propagator, we have also shown that the REG supports the propagation of longitudinal and transverse plasmon modes, whose dispersion relations we have calculated, as well as that of a purely photonic mode with the speed of light in vacuum $c$. The photonic mode corresponds to an EM wave propagating in the gas. The fact that its speed was $c$ implied that the {\it modulus} of the index of refraction was equal to one, and that there was no energy loss to the gas, something novel that we thought should be confirmed and clarified.  In the present article, we  explicitly calculate the energy dissipated to the medium by the EM wave, and show that it vanishes. Furthermore, we interpret this as a natural consequence of the absence of a dissipation mechanism in an ideal gas.

With those two ingredients well established, i.e., simultaneously negative effective permittivity and permeability, and the existence of a photonic mode for which the effective index of refraction $n_{\rm eff}$ has modulus one, we combine them in the discussion of refraction to show that $n_{\rm eff}=-1$, a consequence anticipated in Veselago's work \cite{Veselago}. That completed our proof that the REG is indeed a LHM, something we had suggested previously, but lacked explicit demonstration. Additionally, we illustrate how this affects the reflection and transmission coefficients of the EM wave. We may now finally claim that the REG is a {\sl natural} LHM. 

We remark that by {\sl natural} we mean {\sl not artificially constructed}, unlike the SRR-arrangements previously mentioned. Just as a nonrelativistic free electron gas may be a reasonable approximation to describe the outer electrons of the atoms in a crystal, we expect the REG to be an approximate description of certain regimes of an electron plasma inside a Tokamak \cite{helander2003}, or of certain astrophysical scenarios, such as superdense electron-positron plasmas in Gamma-Ray Burst (GRB) sources, where the density is in the range $n=(10^{30}-10^{37}){\rm cm}^{-3}$ \cite{astrophys, han2012}. In the context of plasma physics, the REG has been the subject of many articles \cite{plasma refs}.

We should also reemphasize that the key ingredient to obtain artificial LHM's was the fabricated enhancement of magnetic responses, since nonrelativistic systems have magnetic responses significantly smaller than electric ones, ultimately because the sources of magnetic fields are current densities, of order $v/c \ll 1$  with respect to the charge densities that generate electric fields. However, in relativistic systems ($v\approx c$) the disparity disappears, making magnetic responses comparable to electric ones. That is why we have resorted to the REG. 

This article is organized as follows: in section \ref{ER}, we describe the theoretical procedure used to obtain the constitutive relations for the REG, derive the effective electric permittivity and effective magnetic permeability that appear in Maxwell's equations, and show that, in the long-wavelength limit, they are both negative below the {\it same} threshold frequency; in section \ref{PROP},  we derive the electromagnetic energy-momentum tensor to show that the energy dissipated by the medium {\it vanishes} throughout the propagation of the photonic mode; in section \ref{RHMLHM}, we show that the REG has an effective index of refraction $n_{\rm eff}= -1$, by discuss refraction and Snell's law; in section \ref{RT}, we present reflection and transmission coefficients for a wave incident from vacuum; finally, we present our conclusions in section \ref{conclusions}.

\section{Veselago Effective responses}
\label{ER}

The REG is a plasma of electrons which, in the absence of interactions, obeys a Fermi-Dirac distribution at very high densities, or very high temperatures, or both. Under those circunstances, either the Fermi energy of the system, or its thermal energy, or both, will be much greater than the electron rest mass, so that many electrons will have relativistic speeds.

The REG has been recently studied \cite{AragaoPRDS2016, ReisAdP2018} by means of a semiclassical treatment of QED at finite temperature and charge density, restricted to a RPA approximation and linear response, in which the EM field was taken as classical, whereas the electrons were subject to a full quantum treatment, just as in the nonrelativististic case that led to the celebrated Lindhard expressions \cite{Lindhard1954}.

The Lagrangean density for the EM field, in Euclidean metric, was shown to be \cite{ReisAdP2018}
\bea
\label{eq:lag}
&& \mathcal{L}_E = \frac{1}{2}\int \frac{d^4q}{(2\pi)^4}{A}_\mu(q){\Gamma}_{\mu\nu}(q){A}_\nu(-q), \\
&& { \Gamma}_{\mu \nu} = q^2 \delta_{\mu \nu} - (1 - \frac{1}{\lambda}) q_\mu q_\nu - {\Pi}_{\mu \nu}.
\label{Gamma}
\eea
$\lambda$ is a gauge parameter, and ${\Pi}_{\mu \nu}$ is the polarization tensor of QED, shown in Appendix \ref{App1}. 

Defining a tensor  ${P}_{\mu\nu}$ whose components comprise the polarization vector $\vec P$, ${P}_{4j}= iP^j$, and the magnetization vector $\vec M$, ${P}_{ij}= - \epsilon_{ijk} M^k$, just as the tensor $ {F}_ {\mu\nu}$ comprises the electric field $\vec{E}$, ${F}_{4j}= iE^j$, and the magnetic field $\vec B$, ${F}_{ij}= \epsilon_{ijk} B^k$, we have obtained \cite{AragaoPRDS2016}
\be
{P}_{\mu\nu} = \frac{\Pi_{\mu\sigma}}{q^2} F_ {\nu\sigma}- \frac{\Pi_{\nu\sigma}}{q^2} F_{\mu\sigma},
\label{Pmunu}
\ee
which led to ${H}_{\mu\nu}\equiv {F}_{\mu\nu}+{P}_{\mu\nu}$, whose components are given in terms of the electric displacement $\vec D$, ${H}_{4j}= iD^j$, and magnetic induction $\vec H$, ${H}_{ij}= - \epsilon_{ijk} H^k$. 

From the calculation of the polarization tensor of QED and the definitions above, we have shown  \cite{AragaoPRDS2016} that the Fourier transformed constitutive equations of the REG are  ${ D}_j=\epsilon_{jk} { E}_k + \tau_{jk}  { B}_k$, $ { H}_j= \nu_{jk}  { B}_k + \tau_{jk}  { E}_k$, with $\nu\equiv \mu^{-1}$. Those tensors may be written as $\epsilon_{jk}= \epsilon \delta_{jk} + \epsilon' \hat{q}_j \hat{q}_k$, $\nu_{jk}= \nu \delta_{jk} + \nu' \hat{q}_j \hat{q}_k$, and $ \tau_{jk}= \tau \epsilon_{jkl} \hat{q}_l$. 
The eigenvalues of $\epsilon_{jk}$ are $\epsilon + \epsilon'$ and $\epsilon$. The eigenvector of $\epsilon+ \epsilon'$ is along $ \hat{q}_k$, thus longitudinal, whereas the two eigenvectors with eigenvalue $\epsilon$ are along directions transverse to $ \hat{q}_k$. The same holds for $\nu_{jk}$, with eigenvalues $\nu + \nu '$ and $\nu$, whereas the tensor $\tau_{jk}$ is clearly transverse.

All the electromagnetic responses that appear in the constitutive relations were calculated from the polarization tensor of QED at finite temperature and density. In particular, their long-wavelength limit, $|\vec{q}|\rightarrow 0$, where the wavelength of the radiation is much larger than the Compton wavelength of the electrons, was explicitly obtained \cite{AragaoPRDS2016} (see Appendix \ref{App1}). 

However, those responses are NOT the ones that appear in Maxwell's equations. In order to find the exact analog of the situation proposed by Veselago \cite{Veselago}, we need the effective electric permittivity and effective magnetic permeability that emerge in the two Maxwell equations with sources. To obtain them, we start with the polarization and magnetization written as 
\be
\label{Polbianiso}
{P}_i=(\epsilon-1){ E}^T_i+(\epsilon_L-1){ E}^L_i-\tau\epsilon_{ijk}\hat{q}_j{ B}^T_k,
\ee
\be
\label{Magbianiso}
{M}_i=(1-\nu){ B}^T_i+(1-\nu_L){ B}^L_i+\tau\epsilon_{ijk}\hat{q}_j{ E}^T_k,
\ee
where the superscripts refer to longitudinal and transverse. Maxwell's equations are
\be
\label{ME1}
\epsilon_{ijk}q_j{ E}_k=\omega{ B}_i,
\ee
\be
\label{ME2}
\epsilon_{ijk}q_j{ H}_k=-\omega{ D}_i.
\ee
For an EM wave, the longitudinal components of the fields ${ B}^L={ E}^L=0$ vanish in eqs.($\ref{Polbianiso}$) and ($\ref{Magbianiso}$), so that the constitutive equations read
\be
\label{D}
{ D}_i=\epsilon{ E}^T_i-\tau\epsilon_{ijk}\hat{q}_j{ B}^T_k,
\ee
\be
\label{H}
{H}_i=\nu {B}^T_i-\tau\epsilon_{ijk}\hat{q}_j {E}^T_k.
\ee
Substituting them into Maxwell's equations yields
\be
\label{ME1_LHM}
	\vec{q}\wedge\vec{E}^T=\omega\left(\frac{\mu|\vec{ q}|}{|\vec{ q}|-\omega\mu\tau}\right)\vec{H}^T,
\ee
\be
\label{ME2_LHM}
\vec{ q}\wedge\vec{ H}^T=-\omega\left(\epsilon+\frac{|\vec{ q}|}{\omega}\tau\right)\vec{ E}^T.
\ee
Comparing with Veselago's work  \cite{Veselago}, we identify the effective responses of the REG as
\be
\label{mueff}
	\mu_{\rm eff}=\frac{\mu|\vec{ q}|}{|\vec{ q}|-\omega\mu\tau}
\ee
\be
\label{epsiloneff}
\epsilon_{\rm eff}=\epsilon+\frac{|\vec{ q}|}{\omega}\tau
\ee
Note that, when $\tau=0$, we recover $\mu_{\rm eff}=\mu$ and $\epsilon_{\rm eff}=\epsilon$. Defining $|n_{\rm eff}| \equiv |\vec{ q}|/\omega$, $n^2_{\rm eff}=\mu_{\rm eff}\epsilon_{\rm eff}$. The LHM behavior will occur whenever $\epsilon_{\rm eff}$ and $\mu_{\rm eff}$ are both negative. 

\begin{figure}[h!]
	\begin{center}
		\includegraphics[width=80mm]{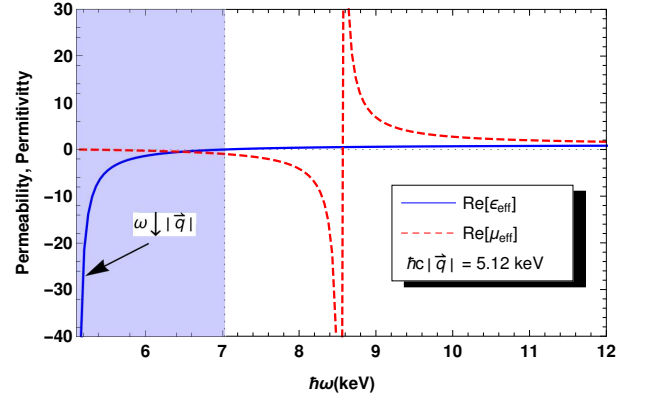}
		\caption{\footnotesize Real parts of $\epsilon_{\rm eff}$ (solid line) and $\mu_{\rm eff}$ (dashed line) as a function of the $\hslash\omega$, and  $\hslash c |\vec{q}|=5.12 {\rm k eV}$, where $|\vec{q}|$ is the wavevector. When $\omega\downarrow |\vec{ q}|$, $\epsilon_{\rm eff}\rightarrow -\infty$. The shaded area correponds to the region where both $\epsilon_{\rm eff}$ and $\mu_{\rm eff}$ are simultaneously negative. Results were obtained  for $T = 0$ and electron gas density $\eta = 1.76\times10^{28}cm^{-3}$. }
		\label{fig1}
	\end{center}
\end{figure}

Fig.\ref{fig1} shows the behavior of the real parts of $\epsilon_{\rm eff}$ and $\mu_{\rm eff}$ at $T=0$ for $|\vec{q}|=5.1 {\rm keV}/\hslash c$. Both effective responses become negative in the shaded region, exhibiting the LHM behavior. In this region, the imaginary parts ${\rm Im}\epsilon_{\rm eff}={\rm Im}\mu_{\rm eff}=0$ are null, so that wave propagation occurs without energy dissipation, something we will confirm by explicit calculation in the next Section.

Using the results of Appendix \ref{App1}, we may write the effective responses, eqs. (\ref{mueff}) and (\ref{epsiloneff}), in terms of the scalar functions $\mathcal{A}^*$ and $\mathcal{B}^*$, and neglect the vacuum contribution, $\mathcal{C}^*<<1$, to derive 
\be
\epsilon_{\rm eff}=1+\mathcal{A}^*-\frac{\omega^2}{|\vec{ q}|^2}\mathcal{B}^*,
\ee
\be
\mu_{\rm eff}=\left(1+\mathcal{A}^*-\frac{\omega^2}{|\vec{ q}|^2}\mathcal{B}^*\right)^{-1}.
\ee
So, we obtain $\epsilon_{\rm eff}=\mu_{\rm eff}^{-1}=\nu_{\rm eff}$, thus $n_{\rm eff}^2=1$. This again implies that an EM wave will propagate with the speed of light in vacuum.

From Appendix \ref{App2}, eqs. (\ref{e20}) and (\ref{e21}), with $\mathcal{C}^*=0$, we derive
\be
\epsilon_{\rm eff}=\nu_{\rm eff}= \nu + \frac{\omega^2}{|\vec{q}|^2+\omega^2} (\epsilon - \nu).
\ee
In the long-wavelength limit $|\vec{q}|\rightarrow 0$, we may use the Drude expressions in eqs.(\ref{Drudee}) and (\ref{Druden}), neglecting small corrections of $\mathcal{O}(\alpha)$, to find
\be
\epsilon_{\rm eff}=\nu_{\rm eff}=\nu= \left(1- \frac{\omega_e^2}{\omega^2}\right).
\ee
Therefore, in the long-wavelength limit, the effective responses will have the exact same Drude behavior, both being negative below the longitudinal electric plasmon frequency, to appear in the next Section. Those simultaneously negative responses characterize the gas as a LHM, as we shall show. In summary, for Fermi gases, the LHM behavior is a characteristic that appears only in a relativistic context.

\section{Propagation without loss}
\label{PROP}

A complete discussion of the modes that propagate in the REG was given in \cite{ReisAdP2018}, where analytic results at $T=0$ and numerical results for $T\neq 0$ were used to compute decay constants and dispersion relations for both longitudinal and transverse plasmons. In order to obtain the collective modes of oscillation, we computed how the medium affects the photon propagator in the REG. The inverse of the quadratic kernel ${ \Gamma}_{\mu \nu}$ in eq.(\ref{Gamma}) gives the photon propagator
\be
{ \Gamma}_{\mu\nu}^{-1}=\frac{\mathcal{P}^L_{\mu\nu}}{-q^2\epsilon_L}+\frac{\mathcal{P}^T_{\mu\nu}}{-q^2(\nu_L+1)}+\frac{\lambda}{q^2}\frac{q_\mu q_\nu}{q^2},
\ee
with the projectors $\mathcal{P}_{\mu\nu}\equiv \mathcal{P}_{\mu\nu}^L+\mathcal{P}_{\mu\nu}^T=\delta_{\mu\nu}-q_\mu q_\nu/q^2$, $\mathcal{P}^T_{ij}=\delta_{ij}-\hat{q}_i\hat{q}_j$, and $\mathcal{P}^T_{44}=\mathcal{P}^T_{4i}=0$; $\lambda$ is the gauge parameter. The poles of the photon propagator correspond to collective excitations and yield their dispersion relations.

In the longitudinal propagator, there is a pole whenever the longitudinal electric permittivity vanishes $\epsilon_L(\omega,\vec{q})=0$, which leads to the dispersion relation of the longitudinal plasmon collective excitation. For $\epsilon_L(\omega,\vec{q})$ nonzero, Maxwell's equations lead to transverse fields, $\vec{q}\cdot\vec{E}=0$, which means that the pole $q^2=0$ in the longitudinal propagator is not realized in this case.

The transverse propagator has poles whenever the inverse of the magnetic permeability becomes $\mu^{-1}_L(\omega,\vec{q})\equiv \nu_L(\omega,\vec{q})=-1$. They correspond to collective oscillations of the current density. There is another transverse mode of propagation in the REG  whenever $q^2=\omega^2-|\vec{q}|^2=0$, corresponding to a photonic mode that propagates with the speed of light $c$ in vacuum. 

Our main interest here will be the photonic mode, which corresponds to an EM wave that travels through the REG as if it were in vacuum. We will show by explicit computation that no energy is dissipated into the medium by the EM wave. We start by constructing the energy-momentum tensor $T_{\mu\lambda}$
\be
T_{\mu\lambda}=F_{\mu\nu}F_{\lambda\nu}-\delta_{\mu\lambda}F^2/4,
\ee
which satisfies
\be
\label{tensorEM}
\partial_\mu T_{\mu\lambda}=-J_{\mu}F_{\mu\lambda},
\ee
with $T_{44}=-u$, $u$ being the energy density , and $T_{j4}=iS_j$, $\vec{ S}$ being the Poynting vector. 

The total current $J_\mu$ is the sum of the free and induced current contributions, $J_\mu=J^{F}_\mu+J^I_\mu$. The current $J^{I}_{\nu}$ induced in the REG by the external EM field is related to the polarization $P_{\mu\nu}$ by $\partial_\mu P_{\mu\nu}=-J^I_\nu$. Since $J^I_4=i\vec{\nabla}\cdot\vec{ P}$ and $J^{I}_j=-\partial_tP_j-(\vec{\nabla}\wedge\vec{M})_j$, eq.(\ref{tensorEM}) may be expressed in Minkowski space, using $J^M_0=iJ^E_4,\ \vec{J}_M=-\vec{J}_E$ 
\be
\label{Poynting}
\frac{\partial u}{\partial t}+\vec{\nabla}\cdot \vec{S}=-\vec{E}\cdot\left(\vec{J}^{F}+\vec{\nabla}\wedge\vec{M}+\frac{\partial\vec{ P}}{\partial t}\right).
\ee
Eq.(\ref{Poynting}) expresses the Poynting theorem \cite{jackson, landau}, with $u=(\vec{ E}^2+\vec{B}^2)/2$,  $\vec{S}=\vec{ E}\wedge\vec{B}$,  and $\vec{M}$ and $\vec{P}$ the magnetization and polarization vectors, respectively. A simple rewriting leads to
\be
\label{poyntingTheo}
\left[\vec{ E}\cdot\frac{\partial\vec{D}}{\partial t}+\vec{H}\cdot\frac{\partial \vec{B}}{\partial t}\right]+\nabla\cdot(\vec{ E}\wedge\vec{ H})=-\vec{ E}\cdot\vec{ J}^{F}.
\ee

The total energy dissipated may be identified as

\be
\mathcal{E}=\int d^4x\left(\vec{ E}\cdot\frac{\partial\vec{ D}}{\partial t}+\vec{ H}\cdot\frac{\partial\vec{ B}}{\partial t}\right).
\ee
In terms of the Fourier transforms,
\be
\mathcal{E}=\int \frac{d^4q}{(2\pi)^4 }i\omega\left[{ E}_j(q){ D}_j(-q)+{ H}_j(q){ B}_j(-q)\right].
\ee
Since the fields are real, $E_j(x)=E^*_j(x)$, we have ${ E}_j(-q)={ E}^*_j(q)$. Inserting the constitutive equations of the REG for ${ H}$ and ${ D}$,
\bea
\mathcal{E}&&=\int \frac{d^4q}{(2\pi)^4}(-i\omega)\Big[\epsilon_L|{ E}_L|^2+\epsilon|{ E}^T_j|^2-\nu_L|{ B}_L|^2\nonumber\\
&&-\nu|{ B}^T_j|^2+2\tau\epsilon_{ijk}\hat{q}_k \text{Re}({ E}^*_i{ B}_j)\Big].
\eea
From Maxwell's equations, we have ${ E}_L={ B}_L=0$, and ${ B}_i={ B}^T_i=\epsilon_{ijk}q_j{ E}^T_k/\omega$. Thus, 
\be
\label{energy}
\mathcal{E}=-i\int \frac{d^4q}{(2\pi)^4}\omega\left[\epsilon(q)-\nu(q)\frac{|\vec{ q}|^2}{\omega^2}+2\tau(q)\frac{|\vec{ q}|}{\omega}\right]|{ E}^T_i(q)|^2.
\ee

For the photonic mode, $|\vec{q}|^2-\omega^2=0$, the term inside the brackets in eq.(\ref{energy}) is 
\be
\label{e34}
\epsilon(q)-\nu(q)+2\tau(q).
\ee
Setting $\mathcal{C}^*=0$ in eqs.(\ref{e20})-(\ref{e23}), eq.(\ref{e34}) becomes
\be
\epsilon(q)-\nu(q)+2\tau(q)=\left(1-\frac{\omega}{|\vec{ q}|}\right)^2\mathcal{B}^*.
\ee
Thus, for the EM wave, $|\vec{ q}|=\omega$ implies $\mathcal{E}=0$ in (\ref{energy}). No energy is dissipated.

In conclusion, the photonic mode propagates with the speed of light in vacuum, meaning that there are no losses to the medium, as explicitly verified by our calculation of the vanishing energy dissipation. Physically, this is inevitable, because there is no mechanism for dissipating energy in the REG, as it is an ideal gas. Its particles do not interact with each other, only with the photons of the external radiation, which will only change the momenta of the electrons, without losing energy (elastic collisions). Had we been dealing with atoms, the photons could induce transitions in the energy levels of the electrons, and that would lead to energy loss.

\section{Negative Index of Refraction}
\label{RHMLHM}

\begin{figure}[h!]
	\begin{center}
		\includegraphics[width=80mm]{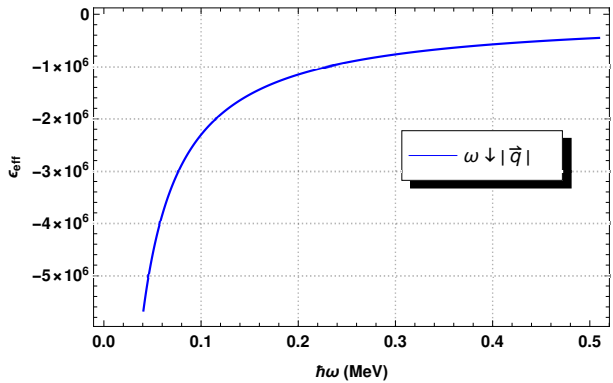}
		\caption{\footnotesize Effective permittivity for values of $\omega\downarrow |\vec{ q}|$ (photonic mode). Calculations were performed at $T=0$ and electron density $\eta= 1.76 \times 10^{28} cm^{-3}$.}
		\label{fig2}
	\end{center}
\end{figure}

For the photonic mode, since $|n_{\rm eff}|=1$, the effective responses in eqs.(\ref{mueff}) and (\ref{epsiloneff}) become $\epsilon_{\rm eff}=\epsilon+\tau$ and $\nu_{\rm eff}=\nu-\tau$, where $\nu_{\rm eff}=\mu_{\rm eff}^{-1}$. Neglecting the vacuum term $\mathcal{C}^*=0$, eqs.(\ref{e20}), (\ref{e21}) and (\ref{e23}) become
\be
\epsilon=1+\mathcal{A}^*,
\ee
\be
\nu=1+\mathcal{A}^*-2\mathcal{B}^*,
\ee
\be
\tau=-\mathcal{B}^*,
\ee
leading to
\be
\epsilon_{\rm eff}=\nu_{\rm eff}=1+\mathcal{A}^*-\mathcal{B}^*.
\ee
The longitudinal responses for the photonic mode are
\be
\epsilon_L=\epsilon+\epsilon'=1,
\ee
\be
\nu_L=\nu+\nu'=1+2(\mathcal{A}^*-\mathcal{B}^*),
\ee
leading to
\be
\epsilon_{\rm eff}=\nu_{\rm eff}=\frac{\nu_L+1}{2}.
\ee
This becomes negative for $\nu_L<-1$ and, then,
\be
\epsilon_{\rm eff}=\nu_{\rm eff}<0.
\ee
A numerical calculation shows that, for the EM wave of the photonic mode $\omega=|\vec{q}|$, $\epsilon_{\rm eff}=\nu_{\rm eff} \rightarrow -\infty$, keeping the product $n^2_{\rm eff}=\mu_{\rm eff}\epsilon_{\rm eff}=1$. That can be seen by taking the limit $\omega \downarrow |\vec{q}|$ from above, since $\omega= \lim_{m_\gamma\rightarrow 0}\sqrt{\vec{q}^2+m_{\gamma}^2}$, $m_\gamma$ being the vanishing photon mass, as shown in Fig.\ref{fig2}. Thus, we have $\epsilon_{\rm eff}<0$ and $\nu_{\rm eff}<0$, for the EM wave.

Maxwell's equations provide the relative orientation of the wave vector $(\hat{q})$ with respect to the electric ($\hat{e}$) and magnetic $(\hat{h})$ fields. For the usual case where the medium is right-handed, we have $\hat{q}\wedge \hat{e}=+\hat{h}$ and $\hat{q}\wedge\hat{h}=-\hat{e}$. On the other hand, Maxwell's equations (\ref{ME1_LHM}) and (\ref{ME2_LHM}), for a LHM with $\mu_{\rm eff}<0$ and $\epsilon_{\rm eff}<0$, yield $\hat{q}\wedge\hat{e}=-\hat{h}$ and $\hat{q}\wedge\hat{h}=+\hat{e}$. From previous considerations, the REG is a LHM. Then,
\be
\label{LHMvectors}
	\hat{e}\wedge\hat{h}=-\hat{q}.
\ee
From the definition of the Poynting vector, $\vec{S}\equiv \vec{ E}\wedge\vec{ H}$, the vector $\hat{s}=\hat{e}\wedge\hat{h}$, is opposite to its wavevector by eq.(\ref{LHMvectors}).

Let us analyze the refraction of light into a LHM. Maxwell's equations lead to boundary conditions on the interface ($z=0$) delimited by the two media
\bea
\label{boundary1}
&& E_{(xy)a}=E_{(xy)b}, \,\,\,\,\,\,\,\, H_{(xy)a}=H_{(xy)b},  \\
&& \epsilon_1 E_{(z)a}=\epsilon_{\rm eff}E_{(z)b}, \,\,\,\,\,\,\,\, \mu_1 H_{(z)a}=\mu_{\rm eff}H_{(z)b}, 
\label{boundary2}
\eea
where $\epsilon_1$ and $\mu_1$ correspond to the responses of a RHM. 

\begin{figure}[h!]
		\begin{center}
		\includegraphics[width=80mm]{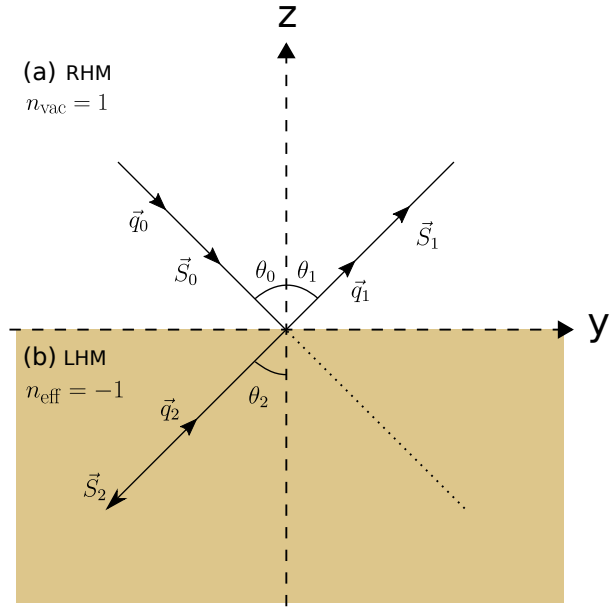}
		\caption{\footnotesize Illustration of the two media with refractive index (a) $n_{\rm vac}=1$  and (b) $n_{\rm eff}=-1$ . When the EM wave enters the LHM (b), the Poynting vector $\vec{ S}_2$  is opposite to $\vec{q}_2$, and the refracted angle is $\theta_2$.}
		\label{fig3}
	\end{center}
\end{figure}

As Veselago argued, the boundary conditions must be satisfied regardless of the relative rightness of the media. Thus, because of the continuity of the tangential component of $\vec{q}$, an incident wave with Poynting vector $\vec{ S}_{\rm 0}$ and wavevector $\vec{ q}_{\rm 0}$ from medium (a) has two possibilities to refract into medium (b), depicted in Fig.\ref{fig3}, depending on the rightness of the vectors fields. If $\epsilon_{\rm eff}$ and $\mu_{\rm eff}$ are simultaneously negative in medium (b), the only way to refract is with an angle $\theta_{\rm 2}$ opposite to the one for a RHM. Since the radiation flows along the Poynting vector $\vec{ S}_{\rm 2}$, which is antiparallel to wavevector $\vec{ q}_{\rm 2}$, Snell's law implies that, for the pure photonic mode, $\omega=|\vec{q}|$, the index of refraction is $n_{\rm eff}=-1$, which completes our proof.

Indeed, let us define the wavevector of the radiation incident from the RHM as $\vec{q_0}$, that of the reflected wave as $\vec{q_1}$, and that of  the wave refracted in the LHM as $\vec{q}_2 $. Then, from the continuity of the tangential components, the boundary conditions on (\ref{boundary1}) and $(\ref{boundary2})$ lead to $q_{0y}=q_{1y}=+q_{2y}$. In the LHM, the tangential component $q_{2y}$ has positive sign. However, its Poynting vector $\vec{ S}_{2y}$ has a minus sign, $S_{0y}=-S_{2y}$. 
Since, $|\vec{ q}_0|=|\vec{ q}_1|=\omega |n_1|$, where $n_1$ is the refractive index of medium $(a)$,  following standard  calculations \cite{landau}, for transparent media $\theta_0=\theta_1$, and we obtain Snell's law
\be
p_1\ n_1\sin\theta_0=p_2\ n_2\sin\theta_2,
\ee
where for a RHM, $p_1=+1$, and for a LHM, $p_2=-1$, with $n_1>0$, and $n_2<0$.

We have thus shown that the only way that the EM wave propagates in the REG is with $n_{\rm eff}=-1$ in the photonic mode $\omega=|\vec{q}|$. This result corrects a misunderstanding in Ref.\cite{ernesto2020}, which misused the concept of index of refraction for the electric plasmon oscillation. The concept of index of refraction can only be applied to a propagating electromagnetic wave, not to a plasmon excitation, which is an oscillation of the electric charge density in the medium.

\section{Reflection and Transmission Coefficients}
\label{RT}

The electric and magnetic fields in a LHM may be obtained from Maxwell's equation. We shall use the boundary conditions given by Eqs.(\ref{boundary1}) and (\ref{boundary2}) in order to find the reflection and transmission coefficients for an incoming wave propagating from medium $(a)$(RHM) to medium $(b)$ (LHM), as shown in Fig.\ref{fig4}. Let us start with the case where the electric field only has the component $\vec{ E}_i=E_i\hat{x}$ (TE-mode).  From (\ref{ME1_LHM}) we obtain
\be
\label{ME4}
H_y=\left(\frac{1}{\omega\mu_{\rm eff}}\right)q_z E_x.
\ee
 Therefore, continuity of tangential components of $\vec{ E}$ and $\vec{ H}$ leads to
 \be
 \label{E&H}
 E_{0}+E_{1}=E_{2}, \, \, \, \, \, \ H_{o}+H_{1}=H_{2},
 \ee
where $E_{0x}=E_0$, $E_{1x}=E_1$ and $E_{2x}=E_2$.  Noting that $q_{1z}=-q_{0z}$, after some algebraic manipulations and using Snell's law for transparent media \cite{landau}, we find
\be
\label{E1}
E_1=\frac{\alpha-\beta}{\alpha+\beta}E_0,
\ee 
\be
\label{E2}
E_2=\frac{2\alpha}{\alpha+\beta}E_0.
\ee
where, $\alpha=|\mu_{\rm eff}|\sin\theta_2\cos\theta_0$, and $\beta=\mu_1\sin\theta_0\cos\theta_2$. These results allow us to compute the reflection and transmission coefficients for the TE-incident wave. For that, we need to compute the Poynting vector $ S_i=\epsilon_{ijk}E_j H_k$. For a LHM, using Maxwell's equations (\ref{ME1_LHM}) and (\ref{ME2_LHM}),
\be
{S}_i=-\sqrt{\frac{\epsilon_{\rm eff}}{\mu_{\rm eff}}}{ E}_j E_j\hat{q}_i.
\ee
where we identified $\hat{q}_i|\vec{q}|/\omega=-|n_{\rm eff}|\hat{q}_i$. 

The reflection coefficient $R = \langle { S}_1 \rangle/\langle { S}_0\rangle$ may be computed as the ratio between the average reflected flux $\langle { S}_1 \rangle$ and the average incident flux $\langle { S}_0 \rangle$. For the transmission coefficient $T = \langle { S}_2 \rangle/\langle { S}_0\rangle $, where $\langle { S}_2 \rangle$ is the average transmitted flux. Assuming  $\langle { S}_1 \rangle$ in the same direction as the wavevector, as we are in a right-handed medium, $\theta_0=\theta_1$, and considering medium (a)  $(n=1)$ and medium (b) $(n_{\rm eff}=-1)$ transparent, from Eqs.(\ref{E1}) and (\ref{E2}) we obtain

\be
\label{reflec}
R=\frac{(\alpha-\beta)^2}{(\alpha+\beta)^2},
\ee

\be
\label{transm}
T=\frac{4\alpha\beta}{(\alpha+\beta)^2},
\ee
which give the reflection and transmission coefficient for a TE-incident wave.  Note that $R+T=1$ as expected, since no energy is absorbed by the LHM, as we have seen before.

Eqs.(\ref{reflec}) and (\ref{transm}) can be further simplified for the photonic mode, $\omega=|\vec{ q}|$, where $\theta_0=\theta_2$, as shown in Fig.\ref{fig4}. We find
\be
\label{reflec2}
R=\left(\frac{|\mu_{\rm eff}|+\mu_1}{|\mu_{\rm eff}|-\mu_1}\right)^2,
\ee

\be
\label{transm2}
T=\frac{4|\mu_{\rm eff}|\mu_1}{\left(|\mu_{\rm eff}|+\mu_1\right)^2}.
\ee
As mentioned before, for the photonic mode $\omega=|\vec{ q}|$, $\epsilon_{\rm eff}=\nu_{\rm eff}\rightarrow -\infty$, or $\mu_{\rm eff}\rightarrow 0$. We then obtain the coefficients of reflection  $R=1$, and transmission $T=0$, for that mode. This implies that, if the radiation propagates in vacuum $(n=1)$, and the external medium is a LHM $(n_{\rm eff}=-1)$, we obtain  total reflection at the interface for any angle of incidence, suggesting that a LHM can be used as a waveguide with no energy dissipation.

\section{Conclusions}
\label{conclusions}

We have shown, from Maxwell's equations, that the REG has effective permittivity and permeability that are both negative at frequencies below the longitudinal electric plasmon frequency. We conclude that the REG is a {\it natural} realization of a LHM, and this occurs because the gas is relativistic.

We have also confirmed that the photonic mode propagates in the REG with the speed of light in vacuum, without losses, by explicitly computing the energy dissipated in the gas and finding that it vanishes, a consequence of the fact the electrons do not self-interact. The REG is thus completely transparent to the photonic mode.

Finally, we use  Snell's law to argue that the index of refraction for the photonic mode is $n_{\rm eff}=-1$, and explore the implications of this fact for the reflection and transmission coefficients. We suggest that the REG can act as a perfect waveguide, with no energy dissipation.
\\

\noindent
\textbf{Acknowledgments.} The authors would like to thank the Brazilian Agencies CNPq and CAPES for partial financial support.
\\

\noindent
\textbf{Disclosures.} The authors declare no conflicts of interest.

\appendix

\section {The Polarization Tensor of QED}
\label{App1}

The polarization tensor of QED, at finite temperature and finite charge density, may be naturally split it into two parts ${\Pi}_{\mu \nu} = {\Pi}_{\mu \nu}^{(v)} + {\Pi}_{\mu \nu}^{(m)}$.  The vacuum contribution 
\be
{ \Pi}_{\mu \nu}^{(v)}=-(q^2\delta_{\mu \nu}-q_\mu q_\nu)\mathcal{C}(q^2),
\ee
and the medium contribution
\bea
\label{med1}
&& -\frac{{\Pi}_{ij} ^{(m)}}{ q^2 }= \left (\delta_{ij} - \frac{q_iq_j}{|\vec{q}|^2}\right ) {\cal{A}} + \delta_{ij} \frac{q_4^2}{|\vec{q}|^2} {\cal{B}},  \\
&& -\frac{{\Pi}_{44} ^{(m)}}{q^2 } = {\cal B},  \,\,\,\,\,\,\,\,\,  -\frac{{\Pi}_{4i} ^{(m)}}{q^2 } = - \frac{q_4 q_i}{|\vec{q}|^2} {\cal B}.
\label{med2}
\eea

 $\mathcal{A} (q_4, |\vec q|)$ and $\mathcal{B} (q_4, |\vec q|)$  are scalar functions  obtained from the Feynman graph used to calculate $\tilde \Pi_{\mu \nu}$, i. e.,

\begin{equation}
{\cal{A}}= \frac{-e^{2}}{2 \pi^{3} q^{2}} \mathrm{Re} \int \frac{d^3 p}{\omega_p} n_{F} (p) \frac{p.(p+q)}{q^2-2p.q} + \left(1- \frac{3q^2}{2|\vec{q}|^2}\right){\cal{B}},
\label{calA}
\end{equation}
and
\begin{equation}
{\cal{B}}=\frac{-e^{2}}{2 \pi^{3} q^{2}} \mathrm{Re} \int \frac{d^3 p}{\omega_p} n_{F} (p) \frac{p.q - 2p_4(q_4 - p_4)}{q^2-2p.q} ,
\label{calB}
\end{equation}
where $p_4=i\omega_p=i \sqrt{|\vec{p}|^2+ m^2}$ and $n_{F}(p)= (e^{\beta(\omega_p - \xi)} +1)^{-1} +  (e^{\beta(\omega_p + \xi)} +1)^{-1} $.
Expressions \eqref{calA} and \eqref{calB}  may be integrated  over angles. ${\cal{C}}$ is obtained from the vacuum polarization contribution \cite{IZ, Kapusta}. The calculation of the medium contribution to the polarization tensor was performed long ago \cite{AP}, and its asymptotic behavior for long wavelengths appeared in \cite{AragaoPRDS2016}. 

\section {The permittivities and inverse permeabilities}
\label{App2}
The components of the linear-response RPA tensors, $\epsilon_{jk}$, $\nu_{jk}$ and $\tau_{jk}$ are given by

\bea
\label{e20}
\epsilon \!\!&=& \!\!1 + {\cal A^*}  - \frac{q^2}{|\vec{q}|^2}  {\cal B^*}
+ \left ( 2 - \frac{\omega^2}{q^2} \right ) {\cal C^*},
\eea
\\
\bea
\label{e21}
\nu  &=& 1 + {\cal A^*}  - 2 \frac{\omega^2}{|\vec{q}|^2} {\cal B^*} 
+ \left ( 2 + \frac{|\vec{q}|^2}{q^2} \right ) {\cal C^*},
\eea
\bea
\label{e22}
\epsilon^{\prime} &=& - {\nu^{\prime}}  = - \left [ {\cal A^*} - \frac{|\vec{q}|^2}{q^2} {\cal C^*}\right ]  ,
\eea
\bea
\label{e23}
\tau  \! &=& \! - \frac{\omega}{|\vec{q}|} \! \! \left [ \!-  \frac{|\vec{q}|^2}{q^2} {\cal C^* + {\cal B^*} }  \right ] \! \!.
\eea
The asterisk denotes continuation to Minkowski space.

For $T=0$, for instance, and in the long wavelength limit, the electric permittivity $\epsilon$ and the inverse of magnetic permeability $\nu$ were shown to have Drude-type responses \cite{AragaoPRDS2016}.
\be
\label{Drudee}
\epsilon=1-\frac{\omega^2_e}{\omega^2}+\frac{e^2}{3\pi^2}g_e\left(\frac{\xi}{m}\right)+\mathcal{O}\left(\frac{\omega^2}{4m^2}\right),
\ee
\be
\label{Druden}
\nu=1-\frac{\omega^2_m}{\omega^2}+\frac{5e^2}{6\pi^2}g_m\left(\frac{\xi}{m}\right)+\mathcal{O}\left(\frac{\omega^2}{4m^2}\right),
\ee
where $\omega_m^2=2\omega_e^2$  and $\omega_e=e^2\eta/m$ is the longitudinal (electric) plasmon frequency; $\eta$ is the electron gas density; $g_e(\xi/m)$ and $g_m(\xi/m)$ are $\mathcal{O}(\alpha)$ small corrections; and $\xi$ is the chemical potential at $T=0$. 

The Drude-type expressions above imply that electric and magnetic responses are simultaneously negative for small frequencies $\omega$. This is only due to the medium contribution, since the vacuum contribution is of order $(\omega^2/4m^2)$, and does not exhibit any such behavior. Taking the long-wavelenght limit after the nonrelativistic limit of the system yields a Drude expression only for $\epsilon$, but not for $\nu$, as discussed in \cite{AragaoPRDS2016}. For the longitudinal responses, one obtains
\bea
\label{long1}
&& \epsilon_L = \epsilon+\epsilon'= 1+{\cal C}^\ast + \left(1-\frac{\omega^2}{|\vec{q}|^2}\right) {\cal B}^\ast, \\
\label{long2}
&& \nu_L = \nu+\nu'=1+2{\cal C} ^\ast+ 2{\cal A}^\ast -2 \frac{\omega^2}{|\vec{q}|^2} {\cal B}^\ast.
\eea


\begin{thebibliography}{99}


\bibitem{Veselago} V. G. Veselago, Sov. Phys. Usp. \textbf{10}, 509 (1968).

\bibitem{Pendry} J. Pendry, Opt. Express \textbf{11}, 639 (2003).


\bibitem{Smith}  D. R. Smith, Willie J. Padilla, D. C. Vier, S. C. Nemat-Nasser, and S. Schultz, Phys. Rev. Lett. \textbf{84}, 4184 (2000).

\bibitem{AragaoPRDS2016} C. A. A. de Carvalho, Phys. Rev. D \textbf{93}, 105005 (2016).

\bibitem{ReyezEPL2016} E. Reyes-G\'omez, L. E. Oliveira, and C. A. A. de Carvalho, Europhys. Lett. \textbf{114}, 17009 (2016).

\bibitem{JPP} C. A. A. de Carvalho and D. M. Reis, J. Plasma Phys. \textbf{84}, 905840112 (2018). 

\bibitem{ReisAdP2018} D. M. Reis, E. Reyes-G\'omez, L. E. Oliveira, and C. A. A. de Carvalho, Annalen der Physik \textbf{530}, 1700443 (2018).


\bibitem{helander2003} P. Helander and D. J. Ward, Phys. Rev. Lett, \textbf{90}, 135004 (2003).
 
\bibitem{astrophys} V. I. Berezhiani, N. L. Shatashvili, and N. L. Tsintsadze, Phys. Scr.
\textbf{90}, 068005 (2015).

\bibitem{han2012}  W. B. Han,  R. Ruffini,  and S. S. Xue,  Phys. Rev. D
\textbf{86}, 084004 (2012).

\bibitem{plasma refs} J. McOrist, D. B.  Melrose, and J. I. Weise, J. Plasma Phys. \textbf{73},  495 (2007);  D. B. Melrose and L. M. Hayes, Austral. J. Phys. \textbf{37}, 639 (1984); B. Jancovici, Il Nuovo Cimento \textbf{25} (2), 428 (1962).

\bibitem{Lindhard1954} J. Lindhard, K. Dan. Vidensk. Selsk., Mat. Fys. Medd. \textbf{28}, 8 (1954).

\bibitem{jackson} J. D. Jackson, {\it Classical Electrodynamics} (Wiley, New
York, 1998)
\bibitem{landau} L. Landau and E. Lifshitz, {\it Electrodynamics of Continuous
Media} (Pergamon Press, New York, 1960).

\bibitem{ernesto2020} J. D. Mazo-Vasques, L. M. Hicpaie-Zuluaga, and E. Reyes-G\'omez, J. Opt. Soc. Am. B \textbf{37}, 211 (2020).

\bibitem{IZ}  C. Itzykson and J. B. Zuber, {\it Quantum Field Theory} (McGraw-Hill, New-York, 1980).

\bibitem{Kapusta} J. I. Kapusta and C. Gale, {\it Finite-Temperature Field Theory} (Cambridge University Press, Cambridge, England, 2006).

\bibitem{AP} I. A. Akhiezer  and S. V. Peletminskii, Sov. Phys. JETP \textbf{11}, 1316 (1969); E. S. Fradkin, {\it Proceedings of the P. N. Lebedev Physics Institute}, vol. \textbf{29}, Consultants Bureau (1967).





















 









 

\end{thebibliography}
\end{document}